\documentclass[aps, pra, twocolumn, notitlepage, nofootinbib]{revtex4-1}

\usepackage{graphicx,amsmath,pstricks,subfig,float}

\begin{document}

\title{Estimates for the number of visible galaxy-spanning civilizations and the cosmological expansion of life}

\author{S. Jay Olson}
 \email{stephanolson@boisestate.edu}
 \affiliation{Department of Physics, Boise State University, Boise, Idaho 83725, USA}

\date{\today}

\begin{abstract}

If advanced civilizations appear in the universe with an ability and desire to expand, the entire universe can become saturated with life on a short timescale, even if such expanders appear rarely.  Our presence in an apparently untouched Milky Way thus constrains the appearance rate of galaxy-spanning Kardashev type III (K3) civilizations, if it is assumed that some fraction of K3 civilizations will continue their expansion at intergalactic distances.  We use this constraint to estimate the appearance rate of K3 civilizations for 81 cosmological scenarios by specifying the extent to which humanity is a statistical outlier.  We find that in nearly all plausible scenarios, the distance to the nearest visible K3 is cosmological.  In searches for K3 galaxies where the observable range is limited, we also find that the most likely detections tend to be expanding civilizations who have entered the observable range from farther away.  An observation of K3 clusters is thus more likely than isolated K3 galaxies. 

\end{abstract}

\maketitle

\section{Introduction}

It is difficult to imagine a scientific discovery that would have a more profound impact than direct observational evidence of advanced civilizations engaged in engineering at the scale of entire galaxies -- the so-called Kardashev type III (K3) civilizations~\cite{kardashev1964}.  Not only would such an observation answer age-old questions about intelligent life, but it could also become a source of new information about the limits of technology and physics~\cite{lacki2015}.  Nevertheless, this version of SETI has only recently begun to attract serious attempts at observation~\cite{bradbury2011,carrigan2012,zackrisson2015,wright2014,wright2014b,griffith2015}, with the largest of these searches to date~\cite{griffith2015} sensitive to technology-induced waste heat from $\approx 10^5$ nearby large and dwarf galaxies, and reporting a null result.

Kardashev's scale~\cite{kardashev1964} was advanced in the 1960's with the hope of informing searches of extraterrestrial life, including searches for galaxy-spanning civilizations.  Now, more than 50 years later, there remain very few quantitative tools to estimate the number of K3 civilizations that could be within range of observation.  The problem is particularly acute now, with search results beginning to be reported -- if $n$ nearby galaxies have been examined for K3 civilizations with null result, what is the interpretation?  Should we have expected to see anything in $n$ galaxies to begin with?  Based on what assumptions?  Here we introduce a hypothesis that, if valid, seems to heavily constrain the range of possibilities, allowing us to make quantitative predictions on the most likely type of positive search result:

\begin{quote}
\textbf{Expansion Hypothesis:}  K3 civilizations have, by definition, already exhibited the necessary technology and behavior characteristics required to expand rapidly beyond the boundaries of their home galaxy, and are thus probable to do so.
\end{quote}

By ``probable," we mean a probability that is not suppressed by many orders of magnitude.  For example, we arrive at our conclusions when the probability for a K3 civilization to expand is of order $10^{-1}$, but they would require revision at $10^{-3}$.  By ``expand rapidly" we refer to an intergalactic wave of colonization that expands spherically outward from the home galaxy at a substantial fraction of the speed of light (we consider here $v \geq .1 c$), generating an expanding cluster of K3 galaxies (use of the word ``cluster" in this context does not indicate a gravitationally bound system – merely a localized collection).  This hypothesis is far from self-evident, but we will argue in the next section that it should be a natural default assumption for K3 civilizations.

We use the expansion hypothesis in the following way:  It has recently been shown, in the context of homogeneous cosmology, that aggressively expanding civilizations can rapidly fill the entire universe with life, even if such expanders appear rarely~\cite{olson2014}.  The timescale for this to happen is controlled by the appearance rate and the expansion speed of these civilizations.  The rate at which the universe fills with advanced life in turn influences the time of arrival distribution for naturally appearing life such as humanity, as the number of untouched ``galaxies to arrive in" is cut off.   Thus, if we specify a scenario by the characteristic speed of the expanders, and specify humanity's relative time of arrival (as mean, $2 \sigma$ latecomer, etc.) then we have fixed the scale of the appearance rate for the expanders.  The expansion hypothesis then asserts that the rate of appearance for all K3 civilizations (including non-expanders) should not be orders of magnitude different from this appearance rate, and we are in a position to calculate observational quantities like ``the expected number of K3 civilizations visible within co-moving radius $R$" for that scenario.

Using this approach, we model 81 cosmological scenarios of the type described in~\cite{olson2014}, where expansion speeds range from $.1 c$ to $.9 c$, humanity is regarded as appearing at the mean time of arrival or as a $1 \sigma$ or $2 \sigma$ latecomer, and three different appearance rate functions for intelligent life are utilized (depending on the formation rate of earthlike planets and assumptions regarding galaxywide extinction events that could delay the appearance of advanced life).  In nearly all scenarios, we find that the co-moving observation distance required to see (on average) a single K3 civilization is cosmological, i.e. at least as far as the universe's homogeneity scale of $\approx.25$ Gly, and much farther in many cases.  There also exist a significant number of scenarios in which the average number of visible civilizations is less than one, no matter how far we are able to look -- this happens, for example, in all scenarios we consider where civilizations expand at $.9 c$. The few scenarios that violate this trend, suggesting that many observable K3 civilizations should be found nearby, correspond to cases in which humanity has arrived improbably late \emph{and} the expansion speed seems improbably slow for a K3-capable civilization.  

When observability is limited to a few Gly (due to the practical limitations of a survey), we find that the probability that K3's are within visible range is dominated by cases in which expanders have entered the visible range from farther away.  Assumptions regarding galaxywide catastrophes (due to gamma ray bursts, etc.) and their effects on the appearance rate of life, even unrealistically severe ones, have a modest effect on our conclusions.

This paper is organized in the following way:  Section II is a brief argument for the expansion hypothesis on grounds that are independent from the main analysis of this paper.  Section III is a review of aggressive expansion scenarios, in which the saturation of the universe by advanced life resembles a first-order cosmological phase transition involving spatially random ``nucleation events" followed by spherical expansion.  The model of observability we use is also developed in this section.  Section IV develops the three life appearance rate models we use, which constitute one of the basic inputs of our analysis.  Section V tabulates the results of our 81 scenarios, organized by the assumed relative appearance time of humanity, while section VI contains a discussion of the results and our conclusions. 

\section{An argument for the expansion hypothesis}

Although discussions about the possible behavior of advanced life tend to be crippled by a severe lack of data, in the present context we have the advantage that the expansion hypothesis refers specifically to K3 civilizations, and that carries a number of starting assumptions and implications to work with.  In particular, the following assumptions seem safe:

\begin{enumerate}
\item They have mastered interstellar travel.
\item They are not adverse to large-scale expansion for some fundamental reason.
\item They place some value on utilizing natural resources at great distances from their origin.
\end{enumerate}

A recent analysis has made a strong case that intergalactic travel is essentially no more difficult or expensive than interstellar travel -- it merely takes longer~\cite{armstrong2013}.  It has also been pointed out that high-speed space travel is likely to be the least of the technological hurtles on the path to K3 capability, when one considers the requirements implicit in the engineering of entire solar systems~\cite{griffith2015}.  Our first assumption thus seems to imply that practical intergalactic travel should easily be available to any K3 civilization.  If they have achieved K3 status, then they have the \emph{means} to continuously expand.

The second assumption also seems to generalize immediately from the case of $\approx 10^{11}$ stars (a single galaxy) to intergalactic travel.  Many possible reasons have been proposed on the subject of why an advanced civilization might choose to stay close to their homeworld and focus inward rather than outward~\cite{cirkovic2008,sagan1983}, but K3 civilizations, by definition, have found none of them to be compelling.  If they have achieved K3 status, they cannot be fundamentally inhibited where large-scale expansion is concerned -- they must have expanded exponentially already on the galactic scale~\cite{newman1981}.  Furthermore, a K3 civilization has some \emph{motive} to utilize resources on a grand scale, following assumption number 3 -- they are not merely neutral on the issue of expansion.  If such maximally-advanced civilizations have developed self-replicating spacecraft so that the cost of such a venture is minimal, even the mildest preference for expansion occurring at any one of the $10^{11}$ solar systems is all that will be required.

Our argument for the expansion hypothesis is essentially that K3 civilizations have, by definition, already exhibited all of the technological capability and behavior requirements of an aggressive expander, and in the absence of some powerful, universal, and not-yet-articulated reason to stop (or slow dramatically) at the boundaries of a home galaxy, it could be assumed that a significant fraction of K3's will continue their expansion at intergalactic distances unless constrained by their encounters with other expanding civilizations.  

\section{Aggressive Expansion Scenarios}

An ``aggressive expansion scenario" is a proposed cosmological phenomenon~\cite{olson2014}, whereby a subset of advanced life appears at random throughout the universe and expands in all directions, saturating galaxies and utilizing resources as they go.  Mathematically, the description is almost identical to bubble nucleation and growth in a first-order cosmological phase transition, due to the common elements of spatially random nucleation and spherical expansion.  Depending on where the practical limits of technology lie (in particular, if life is able to accelerate the conversion of mass in the universe into radiation), heat may also be rapidly released in such a scenario, inducing a backreaction on the cosmic scale factor and pushing the phase transition analogy closer still.  

Here, we consider a simplified scenario which does not take into account heat generation or cosmological backreaction.  We also assume that all aggressive expanders will be of the same behavior type, i.e. they all expand with the same velocity $v$ in the local comoving frame, and the expanding spherical front of galaxy colonization leads to observable changes a fixed time $T$ after the front has passed by.  An approximate uniformity of behavior of this kind would be expected if the limits of practical technology induce an attractor state in the development of aggressive expanders throughout the universe, though this is only one possibility.

In such a scenario with uniform expansion behavior, the fraction of the universe that remains unsaturated with life, $g(t)$, can be expressed in Guth-Tye-Weinberg~\cite{guth1980,guth1981} (GTW) form as:
\begin{eqnarray}
g(t) = e^{- \int_0^t f(t') V(t',t) dt'}
\end{eqnarray}
where $f(t)$ is the appearance rate of expanding civilizations per unit co-moving volume, per unit time, and $V(t',t)$ is the volume of space fully saturated with life at time $t$ by a single civilization that began expanding at $t'$.  The next section will focus on models for $f(t)$.  When there is a time delay $T$ between the initial arrival of expanding spacecraft at some point in space and the full saturation of matter there (resulting in observable changes), the volume function is given by:
\begin{eqnarray}
V(t',t) = \frac{4 \pi}{3} \left( \int_{t'}^{t-T} \frac{v \ \theta(t''-t')}{a(t'')} dt''  \right)^3
\end{eqnarray}
where $\theta(t)$ is the Heaviside step function and $a(t)$ is the cosmic scale factor\footnote{$a(t)$ is taken to be a flat FRW solution with $ \Omega_{\Lambda 0} = .683$, $\Omega_{r 0} = 3 \times 10^{-5} $, $\Omega_{m 0} = 1-\Omega_{\Lambda 0} - \Omega_{r 0} $, and $H_0 = .069 \ Gyr^{-1} $, fixing the present age of the universe at $t_{0} \approx 13.75$ Gyr. We work in co-moving coordinates, and use units of Gyr and Gly for dimensions of time and distance.}.  Given a background cosmological solution, then, an aggressive expansion scenario is specified by giving $\left\{v,T,f(t) \right\}$.  For the scenarios we examine, $T$ will play a very minor role in the quantities we calculate and could be set to zero as an additional simplification, but for the sake of completeness we will take $T$ to correspond to an ample galaxy colonization time~\cite{hart1975} of $.01$ Gyr, and leave it constant throughout our analysis.

We now come to a key point of our analysis.  Consider the set of all human-stage civilizations who will ever have appeared in the universe, and who, like humanity, have appeared within a non-K3 galaxy.  Our prior assumption is that we (humanity) are ``typical" within this set, and in particular that our time of arrival is typical within this set.  We also assume that the appearance rate for this set has the same baseline cosmic time dependence as $f(t)$, though the overall proportionality constant could be different by a large (and unknown) constant factor.  The time of arrival distribution for this set, however, must be proportional to the product $g(t) f(t)$, to account for the universe filling up with K3 galaxies, in which no members of humanity’s set may thereafter appear.  In other words, the factor of $g(t)$ changes the distribution of arrival times significantly by cutting off the appearance rate at late times, as the universe fills with advanced life and the possibility of evolving in an empty galaxy abruptly comes to an end. If $g(t) = 1$ forever (i.e. if there are no aggressive expanders), then the distribution and our relative time of arrival are fixed by the assumptions going into the construction of $f(t)$. Including the possibility of aggressive expansion, however, allows more possibilities and in particular we can find (by numerical search) expansion scenario parameters that put humanity at $t_0 = 13.75$Gyr at the mean time of arrival or as a $1σ$ or $2σ$ latecomer. The mean time of arrival, $\mu$, and standard deviation, $\sigma$, are given by:

\begin{eqnarray}
\mu &=& N \int_0^\infty t \, g(t) f(t) \, dt \\
\sigma &=& N \sqrt{\frac{1}{N} \int_0^\infty t^2 \, g(t) f(t) \, dt - \left(\int_0^\infty t \, g(t) f(t) \, dt \right)^2}
\end{eqnarray}
where $N = \left( \int_0^\infty g(t) f(t) \, dt \right)^{-1}$ for normalization.  This normalization factor is the reason the unknown proportionality constant for the appearance rate for human-stage life does not affect the time of arrival distribution – one can see that multiplying $f(t)$ by any constant factor will not change the mean or standard deviation of arrival times.

The most powerful search to date~\cite{griffith2015} (by several orders of magnitude) for K3 civilizations has involved data from full-sky surveys of limited range, so we will be interested in calculating $EV(obs)$, the average number (expected value) of civilizations that are observable out to some co-moving distance $R$.  Here, $R$ is supposed to represents a limit to one's equipment and observation techniques (it can also be interpreted as the time $t_1$ appearing in figure 1, and is connected to $t_0$ through $R = \int_{t_1 + T}^{t_0} \frac{1}{a(t)}\, dt$).  Regions within the past light cone that can produce an observable expanding civilization are illustrated in figure 1:  $A$ corresponds to a region in which any produced K3's are directly observable, while $B$ represents a region in which a civilization could appear and, if expanding aggressively, would arrive within $A$ and saturate galaxies there, making that civilization observable within $R$.  The region $C$ (bounded by a ``past saturation cone") is excluded because any expanders appearing there would by now have fully saturated our own galaxy with advanced life, and this is assumed to be ruled out by observation.  Expanders can thus be seen if they appear in $A \cup B$, while non-expanding K3's can be seen if they appear in $A \cup C$.

\begin{figure}
\centering
\includegraphics[width=.9\linewidth]{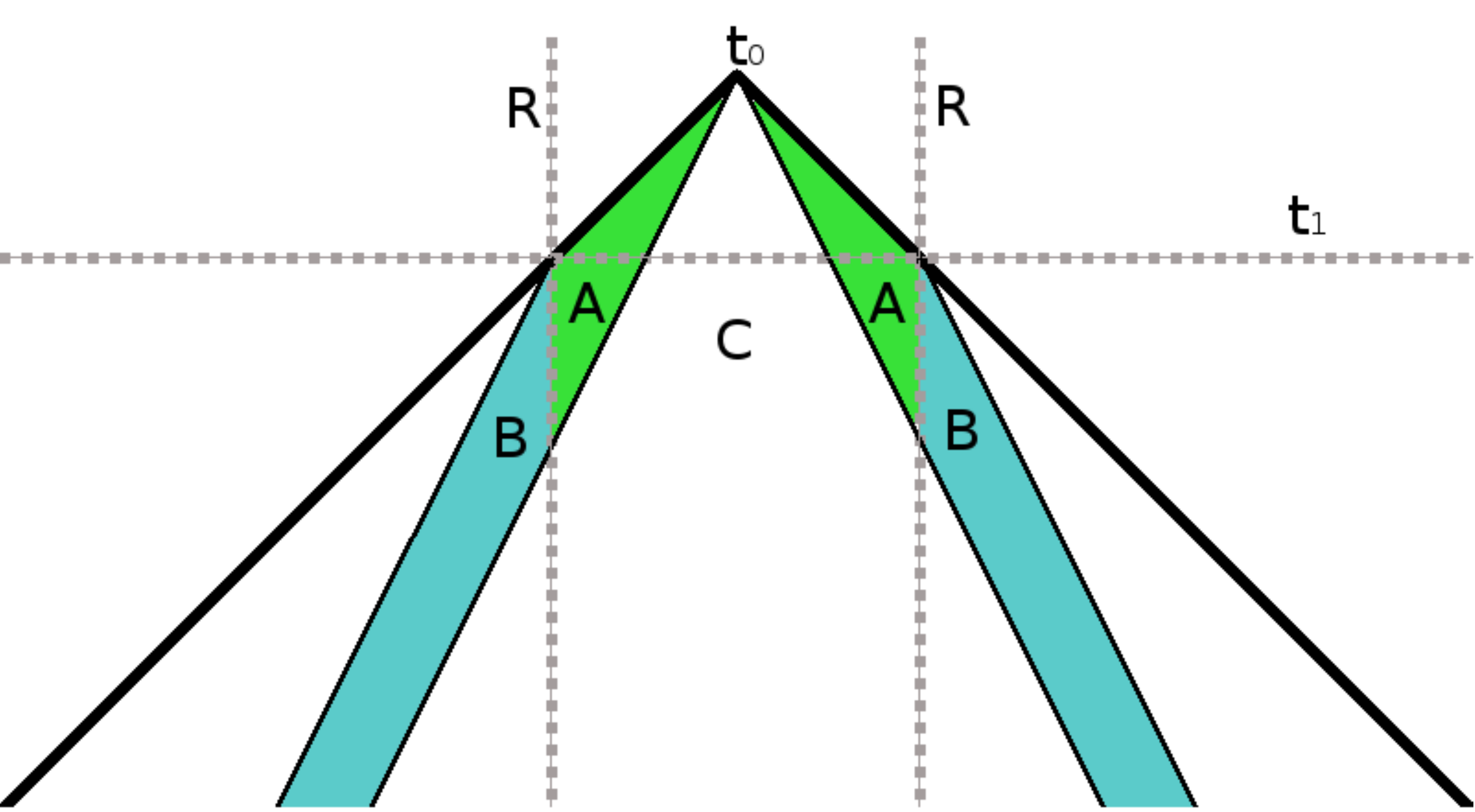}
\caption{The past light cone of an observer at $t_0$ with regions highlighted in which the appearance of an expanding civilization is detectable, under the assumption that only galaxies out to distance $R$ can be directly analyzed for the presence of K3's.  Expanding civilizations appearing in region $B$ are observable because they expand into region $A$.  Region $C$ is presumed to be devoid of expanders because appearing in $C$ would imply that our own galaxy is already fully saturated with maximally advanced life.}
\label{Figure1}
\end{figure}

Considering first only the expanding civilizations, an upper bound for $EV(obs)$ can be expressed by:
\begin{eqnarray}
EV(obs) < \int_{0}^{t_0} f(t) \, \tilde{V}_{R}(t,t_0)  \, dt
\end{eqnarray}
where the volume $\tilde{V}_{R}(t,t_0)$ of region $A \cup B$ at time $t$ is:

\begin{widetext}
\begin{eqnarray}
   \tilde{V}_{R}(t,t_0) = \left\{
     \begin{array}{lcc}
       0 & : &  t > t_0 - T  \\
       \frac{4 \pi}{3} \left( \int_{t +T}^{t_0} \frac{1}{a(t') }\, dt' \right)^{3} - \frac{4 \pi}{3} \left( \int_{t}^{t_0 - T} \frac{v}{a(t')} \, dt' \right)^{3}   & : &  t_0 - T  \geq t \geq t_1 \\
        \frac{4 \pi}{3} \left(R + \int_{t}^{t_1} \frac{v}{a(t')} \, dt' \right)^{3} - \frac{4 \pi}{3} \left( \int_{t}^{t_0 - T} \frac{v}{a(t')} \, dt' \right)^{3}  & : & t_1 > t 
     \end{array}
   \right. 
\end{eqnarray} 
\end{widetext}

This should be regarded as an upper bound because it includes the possibility of ``virtual civilizations" -- expanding civilizations that appear within already-saturated space -- and these should not be counted as independently observable events.  A lower bound on $EV(obs)$ can be expressed as:
\begin{eqnarray}
EV(obs) > \int_{0}^{t_0} g(t) \, f(t) \, \tilde{V}_{R}(t,t_0)  \, dt.
\end{eqnarray}
This is a lower bound because the fraction of unsaturated space in $A$ and $B$ is likely to be higher than $g(t)$, since we have additional knowledge that no expanders from $C$ have saturated any of the space in $A$ and $B$.

For any given expansion scenario (which we will choose by fixing the relative arrival time of humanity, and the velocity of the expanders), we will want to find a characteristic distance $R$ that represents the power of observation required to see an expanding civilization.  We will do this by performing a numerical search of $R$ such that $EV(obs)=1$, referring to the solution as $R_1$.  Because $EV(obs)=1$, the error contributed by virtual civilizations will tend to be small, and we will thus use the upper bound given above as our estimate for $EV(obs)$ in the numerical search of section V.  In this approximation, $R_1$ can also be interpreted as the observation distance at which the probability to see \emph{zero} expanding civilizations is $e^{-1} \approx 37 \% $ (the assumption being that the appearance of K3's is a Poisson process, so in general $P(0) = e^{-EV(obs)}$). 

Although full-sky surveys have been most powerful so far, deep field surveys might also be used, which would correspond to a small angle in the sky, but an unlimited $R$, i.e. a $t_1$ which extends back to the time when $f(t)$ first becomes significantly different from zero.  Thus, for each of the scenarios we will examine, we also estimate $EV(obs)$ (using the upper bound) for the case of an unlimited $R$. 

Finally, we can add non-expanding K3 civilizations to the analysis.  In keeping with the expansion hypothesis, for each scenario considered we will also calculate the expected number of visible but non-expanding K3 galaxies out to $R_1$ under the assumption that the appearance rate for non-expanders is identical to the appearance rate for expanders.  This corresponds to evaluating $\int_{t_1}^{t_0} f(t) \, \hat{V}_{R}(t,t_0)  \, dt$ where $\hat{V}_{R}(t,t_0)$ is the volume associated with region $A \cup C$ from figure 1:
\begin{eqnarray}
   \hat{V}_{R}(t,t_0) = \left\{
     \begin{array}{lcc}
       0 & : &   t > t_0 - T  \\
       \frac{4 \pi}{3} \left( \int_{t +T}^{t_0} \frac{1}{a(t') }\, dt' \right)^{3}   & : &   t_0 - T  \geq t \geq t_1 \\
        \frac{4 \pi}{3} R^{3}  & : &  t_1 > t. 
     \end{array}
   \right.
\end{eqnarray}

\section{Appearance Rate Models}

A basic input of an aggressive expansion scenario is the appearance rate of expanders per unit coordinate volume, per unit cosmic time, $f(t)$.  We will consider three such models for the time-dependence of $f(t)$, leaving the overall proportionality constant as a parameter to be fixed by assumptions on the relative time of arrival of humanity in the next section.  The baseline ``non-catastrophic" model will set the appearance rate at time $t$ to be proportional to the number of earthlike planets formed between $4.5 \, Gyr - 6 \, Gyr$ prior to $t$.  This means that we assume it takes at least 4.5 Gyr for maximally advanced life to appear on a newly formed earthlike planet, and that the window for life to evolve is no more than 6 Gyr.  This assumption is heavily influenced by the successful evolution of intelligence on the Earth, guiding our intuition that conditions should be earthlike, and this assumption could easily be modified if conditions need not be too similar to that of the Earth.  The effect of extending the ``closing of the window" for the evolution of intelligent life is to move back the maximum value of $f(t)$, but would have little effect on our analysis up to the present time, $t_0$.  Adjusting the time until the ``opening of the window," however, will shift the initial rise of $f(t)$ in cosmic time, though we expect that the opening of the window is less likely to vary substantially from our estimate. 

The baseline, non-catastrophic model can be expressed as:
\begin{eqnarray}
f(t) = \alpha \int_{t-6}^{t-4.5} PFR(t') \, dt'
\end{eqnarray}
where the planet formation rate, $PFR(t)$, is modeled by $PFR(t) =N \, M(t) \, SFR(t)$ with $M(t)$ representing an average universe metallicity and $SFR(t)$ the star formation rate of the universe and $N$ a normalization constant.  The overall proportionality constant, $\alpha$, is a free parameter to be fixed by time-of-arrival considerations in the next section. The buildup of metallicity in the universe is, in turn, modeled as an integral over the star formation rate, $M(t) = \int_{0}^{t} SFR(t') \, dt'$, and we express the star formation rate as
\begin{eqnarray}
   SFR(t) = \left\{
     \begin{array}{lcc}
       \frac{t}{3} 10^{t-3} & : &   t < 3  \\
       10^{-\frac{(t-3)}{13.75 - 3}} & : &  t \geq 3 .
     \end{array}
   \right.
\end{eqnarray}
Here, $t$ is in units of Gyr, representing a simple approximation to the SFR data in~\cite{lineweaver2001}.  The overall normalization for $SFR(t)$ and $PFR(t)$ are chosen such that their maximum values are equal to unity.  The choice of normalization is essentially arbitrary in this model, corresponding to a rescaling of $\alpha$.

This model of $PFR(t)$ is a simplified version of Lineweaver's model~\cite{lineweaver2001}, which additionally takes into account a stellar distribution over metallicity.  The resulting $PFR(t)$, plotted in figure 2, mirrors the major features of that model.
\begin{figure}
\centering
\includegraphics[width=0.9\linewidth]{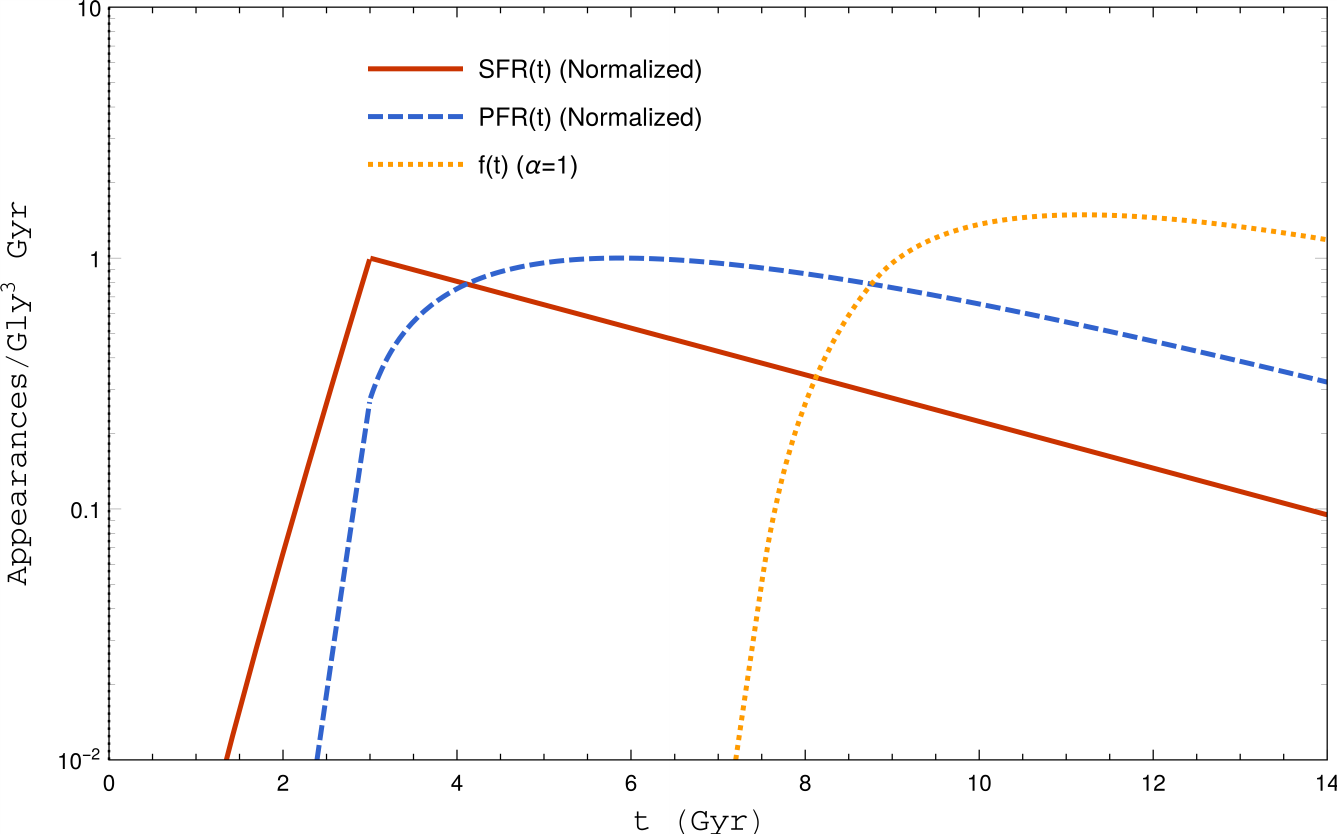}
\caption{Star formation rate $SFR(t)$, planet formation rate $PFR(t)$, and the appearance rate for aggressively expanding life $f(t)$ (for $\alpha=1)$.  $SFR$ and $PFR$ are normalized to a maximum value of unity.  }
\label{Figure2}
\end{figure}

In addition to the baseline model, we introduce two models that include galaxywide extinction events, with a rate that changes as a function of cosmic time.  These models assume that a life-harboring planet will be subject to a high-energy event, nearby gamma ray bursts (GRB's) being the prototype example~\cite{annis1999}, that destroys planetary ozone layers, causing a mass extinction event which sets back the evolution of life by some amount of time.  We will assume that such events are severe if they occur in the final stages of evolution towards intelligence, so that we reduce the pool of potential planets to those which have not seen such an extinction event in the last $.2$ Gyr.  This number was chosen to agree with the estimate given by Annis~\cite{annis1999}, though it may be in the severe range of what is plausible.  The extinction events are modeled as an inhomogeneous Poisson process with intensity $\lambda(t)$ such that $\lambda(t_0)=\frac{1.3}{Gyr}$ -- this value is chosen to be in agreement with a recent analysis~\cite{piran2014} that found the probability of a ``biospherically important event" occurring on the Earth to be $50\%$ in the last $.5$ Gyr.  

The two catastrophic models we present differ in their assumptions regarding the time-dependence of $\lambda(t)$ -- one tracks the observed rate of GRB's in the universe as a function of cosmic time, while the other is an extreme scenario, with past extinction events far more common than suggested by the GRB model.  While we expect the GRB-tracking model to be realistic, the extreme catastrophic model is included to illustrate the extent to which our conclusions change under drastic changes to the life appearance model, and is not intended to be realistic.

To be clear, we are not attempting to model \emph{all} extinction events in this analysis -- only those extinction events whose rate we expect to change strongly as a function of cosmic time.  The rate and effects of local, planetary-scale events are assumed to be approximately equivalent in cosmic time for sufficiently earthlike conditions, and are thus absorbed into the overall proportionality constant $\alpha$.

Modeling these catastrophes amounts to multiplying the baseline appearance rate model for $f(t)$ with the probability that a potential planet has not experienced a catastrophe in the last $.2$ Gyr.  For an inhomogeneous Poisson process with intensity $\lambda(t)$, this probability is $e^{-\int_{t-.2 Gyr}^{t} \lambda(t') \, dt'}$.  The GRB-tracking catastrophic model is given by:

\begin{eqnarray}
\lambda_{GRB}(t) = \frac{1.3}{Gyr} a(t)^{-2.1}
\end{eqnarray}
 where the time-dependence comes from the GRB rate proportional to $(1 + z)^{2.1}$ found by Wanderman and Piran~\cite{wanderman2010}.  The extreme catastrophic model is given by:
\begin{eqnarray}
\lambda_{extreme}(t) = \frac{1.3}{Gyr} e^{-\frac{1}{2}(t - t_0) }
\end{eqnarray}
and the time-dependence is chosen arbitrarily.  The effects of these models on $f(t)$ are plotted in figure 3.

\begin{figure}
\centering
\includegraphics[width=0.9\linewidth]{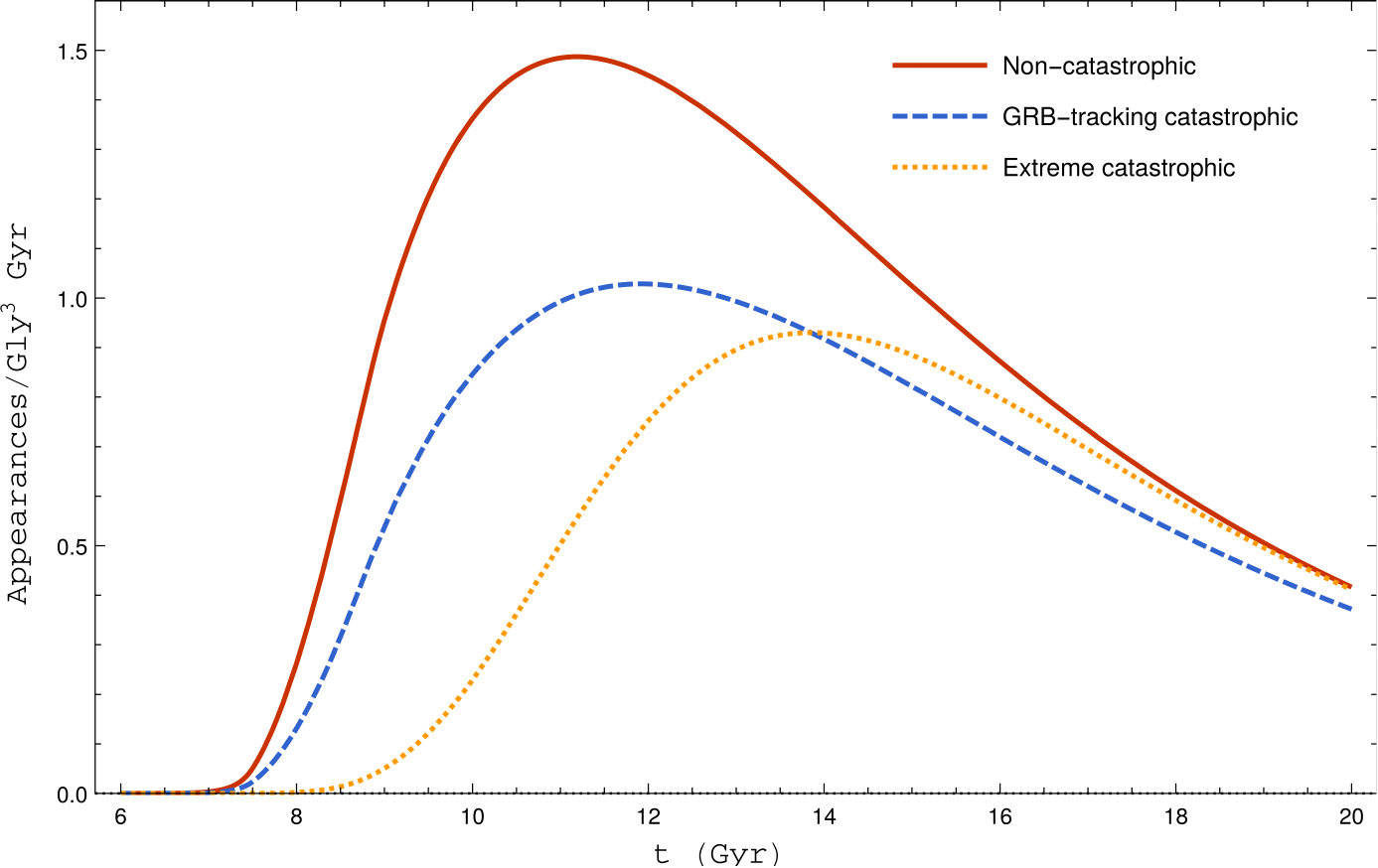}
\caption{ The three appearance rate models for $f(t)$ (for $\alpha = 1$), corresponding to the non-catastrophic baseline model, the GRB-tracking catastrophic model, and the extreme catastrophic model. }
\label{Figure3}
\end{figure}

As mentioned in the previous section, if there are no expanders, then the time-dependence of $f(t)$ alone (and not the proportionality constant $\alpha$) specifies the time of arrival (TOA) distribution of intelligent life.  Table I lists the mean time of arrival, $\mu$, and standard deviation, $\sigma$, for each model.  In all three models, the present time, $t_0 \approx 13.75$ Gyr, is a slightly early but completely unsurprising time of arrival -- this picture will change substantially in the next section when the effect of expanders is included.  In particular, $\mu$ and $\sigma$ will be heavily influenced by the appearance and expansion rates of the aggressive expanders.
\begin{table}[H]
\begin{tabular}{|c||c|c|}
\hline  & $\mu$ (Mean) & $\sigma$ (S.D.) \\ 
\hline Non-Catastrophic & 14.9 Gyr & 5.3 Gyr \\ 
\hline GRB-Tracking Catastrophic & 15.6 Gyr & 5.5 Gyr \\ 
\hline Extreme Catastrophic & 17.1 Gyr & 5.2 Gyr \\ 
\hline 
\end{tabular}
\caption{Mean times of arrival and standard deviation for the three life appearance models, when the effect of aggressive expansion is discounted.  In all models, the appearance time of humanity at $t_0 = 13.75$ Gyr is typical but slightly early.} 
\end{table}

\section{Model Results}

Having described our modeling assumptions and techniques, we are now in a position to numerically examine a range of scenarios, looking for their observational consequence.  We will divide this into three subsections -- one for each assumed time of appearance for humanity, i.e. $t_0 = \mu$,  $t_0 = \mu + \sigma$, or $t_0 = \mu + 2 \sigma$.  To reiterate, $t_0$ remains equal to $13.75$ Gyr in all cases -- it is only the \emph{relative} time of arrival that changes between scenarios (due to the scenario-dependent TOA distribution).  Then, for each appearance time, we will examine three sets of scenarios corresponding to each appearance model (non-catastrophic, GRB-tracking catastrophic, and extreme catastrophic).  Each set then consists of nine scenarios, corresponding to expansion speeds from $.1 c$ to $.9 c$.

Each scenario is obtained through a numerical search of $\alpha$ required to satisfy the time of arrival condition.  After each scenario is obtained, we numerically find and list the values for $R_1$ such that the expected number of observable expanding civilizations is equal to unity, as described in section III (also corresponding to the distance one would have to look to have a probability of $e^{-1} \approx 37 \%$ of \emph{seeing zero expanders}).  The expected number of visible civilizations, $EV(obs)$, for unlimited $R$ is also reported, as will be the expected value of observable non-expanders (isolated K3 galaxies), under the assumption that their appearance rate is identical to that of the expanders (in keeping with the expansion hypothesis).

In a significant number of cases, particularly associated with the mean time of arrival condition, $EV(obs)$ is less than unity for unlimited $R$ (i.e. $R_1$ is undefined), due to the expanders having being made extremely rare to satisfy the time of arrival condition.  These cases will be marked with N/A for the relevant quantities.

\begin{widetext}

\newpage

\subsection{Humanity ($t_0=13.75$ Gyr) at the mean time of arrival}
\subsubsection{Non-catastrophic appearance model}
\begin{table}[H]
\centering
\begin{tabular}{|c||c|c|c|c|c|c|c|c|c|}
\hline Expansion Speed  & $v=.1c$ & $v=.2c$  & $v=.3c$ & $v=.4c$  & $v=.5c$  & $v=.6c$ & $v=.7c$ & $v=.8c$ & $v=.9c$ \\ 
\hline $\alpha$ (appearances/Gly$^3$Gyr) & .019 & .0024 & .00071 & .00030 & .00015 & .000089 & .000056 & .000038 & .000026 \\ 
\hline $R_1$ (Gly) & 1.0 & 2.4 & 5.0 & N/A & N/A & N/A & N/A & N/A & N/A \\ 	
\hline $EV(obs)$ for unlimited $R$ & 35 & 4.3 & 1.2 & .51 & .24 & .13 & .066 & .033 & .013 \\ 
\hline $EV(obs)$ of non-expanders within $R_1$ & .52 & .69 & .95 & N/A & N/A & N/A & N/A & N/A & N/A \\ 
\hline 
\end{tabular} 
\caption{Mean TOA, non-catastrophic appearance model}
\end{table}

\subsubsection{GRB-tracking catastrophic appearance model}
\begin{table}[H]
\centering
\begin{tabular}{|c||c|c|c|c|c|c|c|c|c|}
\hline Expansion Speed  & $v=.1c$ & $v=.2c$  & $v=.3c$ & $v=.4c$  & $v=.5c$  & $v=.6c$ & $v=.7c$ & $v=.8c$ & $v=.9c$ \\ 
\hline $\alpha$ (appearances/Gly$^3$Gyr) & .055 & .0069 & .0020 & .00086 & .00044 & .00025 & .00016 & .00011  & .000075 \\ 
\hline $R_1$ (Gly) & .77 & 1.8 & 3.4 & N/A & N/A & N/A & N/A & N/A & N/A \\ 	
\hline $EV(obs)$ for unlimited $R$ & 57 & 7.1 & 2.1 & .84 & .40 & .21 & .11 & .054 & .021 \\ 
\hline $EV(obs)$ of non-expanders within $R_1$ & .44 & .58 & .79 & N/A & N/A & N/A & N/A & N/A & N/A \\ 
\hline 
\end{tabular} 
\caption{Mean TOA, GRB-tracking catastrophic appearance model}
\end{table}

\subsubsection{Extreme catastrophic appearance model}
\begin{table}[H]
\centering
\begin{tabular}{|c||c|c|c|c|c|c|c|c|c|}
\hline Expansion Speed  & $v=.1c$ & $v=.2c$  & $v=.3c$ & $v=.4c$  & $v=.5c$  & $v=.6c$ & $v=.7c$ & $v=.8c$ & $v=.9c$ \\ 
\hline $\alpha$ (appearances/Gly$^3$Gyr) & .54 & .068 & .020 & .0085 & .0043 & .0025 & .0016 & .0011 & .00075 \\ 
\hline $R_1$ (Gly) & .40 & .92 & 1.7 & 2.9 & N/A & N/A & N/A & N/A & N/A \\ 	
\hline $EV(obs)$ for unlimited $R$ & 110 & 13 & 3.9 & 1.6 & .75 & .39 & .21 & .10 & .040  \\ 
\hline $EV(obs)$ of non-expanders within $R_1$ & .34 & .45 & .61 & .87 & N/A & N/A & N/A & N/A & N/A \\ 
\hline 
\end{tabular} 
\caption{Mean TOA, extreme catastrophic appearance model}
\end{table}

\newpage

\subsection{Humanity ($t_0=13.75$ Gyr) as a $1 \sigma$ latecomer}

\subsubsection{Non-catastrophic appearance model}
\begin{table}[H]
\centering
\begin{tabular}{|c||c|c|c|c|c|c|c|c|c|}
\hline Expansion Speed  & $v=.1c$ & $v=.2c$  & $v=.3c$ & $v=.4c$  & $v=.5c$  & $v=.6c$ & $v=.7c$ & $v=.8c$ & $v=.9c$ \\ 
\hline $\alpha$ (appearances/Gly$^3$Gyr)  & .25 & .031 & .0093 & .0039 & .0020 & .0012 & .00073 & .00049 & .00034 \\ 
\hline $R_1$ (Gly) & .24 & .55 & .94 & 1.5 & 2.3 & 3.6 & N/A & N/A & N/A \\ 	
\hline $EV(obs)$ for unlimited $R$ & 450 & 56 & 16 & 6.6 & 3.2 & 1.6 & .86 & .43 & .17 \\ 
\hline $EV(obs)$ of non-expanders within $R_1$ & .10 & .14 & .20 & .30 & .49 & .85 & N/A & N/A & N/A \\ 
\hline 
\end{tabular} 
\caption{$ \sigma$ latecomer, non-catastrophic appearance model}
\end{table}

\subsubsection{GRB-tracking catastrophic appearance model}
\begin{table}[H]
\centering
\begin{tabular}{|c||c|c|c|c|c|c|c|c|c|}
\hline Expansion Speed  & $v=.1c$ & $v=.2c$  & $v=.3c$ & $v=.4c$  & $v=.5c$  & $v=.6c$ & $v=.7c$ & $v=.8c$ & $v=.9c$ \\ 
\hline $\alpha$ (appearances/Gly$^3$Gyr) & .51 & .063 & .019 & .0079 & .0040 & .0023 & .0015 & .00099 & .00069 \\ 
\hline $R_1$ (Gly) & .21 & .47 & .82 & 1.3 & 2.0 & 3.1 & 7.3 & N/A & N/A \\ 	
\hline $EV(obs)$ for unlimited $R$ & 520 & 65 & 19 & 7.7 & 3.7 & 1.9 & 1.0 & .50 & .19 \\ 
\hline $EV(obs)$ of non-expanders within $R_1$ & .090 & .12 & .18 & .27 & .43 & .74 & 1.5 & N/A & N/A \\ 
\hline 
\end{tabular} 
\caption{$ \sigma$ latecomer, GRB-tracking catastrophic appearance model}
\end{table}

\subsubsection{Extreme catastrophic appearance model}
\begin{table}[H]
\centering
\begin{tabular}{|c||c|c|c|c|c|c|c|c|c|}
\hline Expansion Speed  & $v=.1c$ & $v=.2c$  & $v=.3c$ & $v=.4c$  & $v=.5c$  & $v=.6c$ & $v=.7c$ & $v=.8c$ & $v=.9c$ \\ 
\hline $\alpha$ (appearances/Gly$^3$Gyr)  & 3.7 & .46 & .14 & .057 & .029 & .017 & .011 & .0072 & .0050 \\ 
\hline $R_1$ (Gly) & .12 & .27 & .47 & .73 & 1.1 & 1.7 & 3.0 & N/A & N/A \\ 	
\hline $EV(obs)$ for unlimited $R$ & 720 & 90 & 26 & 11 & 5.1 & 2.6 & 1.4 & .69 & .27 \\ 
\hline $EV(obs)$ of non-expanders within $R_1$ & .070 & .096 & .14 & .21 & .33 & .57 & 1.1 & N/A & N/A \\ 
\hline 
\end{tabular} 
\caption{$ \sigma$ latecomer, extreme catastrophic appearance model}
\end{table}

\newpage
\subsection{Humanity ($t_0=13.75$ Gyr) as a 2$\sigma$ latecomer}
\subsubsection{Non-catastrophic appearance model}
\begin{table}[H]
\centering
\begin{tabular}{|c||c|c|c|c|c|c|c|c|c|}
\hline Expansion Speed  & $v=.1c$ & $v=.2c$  & $v=.3c$ & $v=.4c$  & $v=.5c$  & $v=.6c$ & $v=.7c$ & $v=.8c$ & $v=.9c$ \\ 
\hline $\alpha$ (appearances/Gly$^3$Gyr)  & .92 & .12 & .034 & .014 & .0074 & .0043 & .0027 & .0018 & .0013 \\ 
\hline $R_1$ (Gly)  & .089 & .20 & .35 & .54 & .81 & 1.2 & 1.9 & 3.5 & N/A \\ 
\hline $EV(obs)$ for unlimited $R$ & 1700 & 210 & 60 & 24 & 12 & 6.0 & 3.2 & 1.6 & .61 \\ 
\hline $EV(obs)$ of non-expanders within $R_1$ & .019 & .027 & .040 & .062 & .10 & .19 & .42 & 1.2 & N/A \\ 
\hline 
\end{tabular} 
\caption{$2 \sigma$ latecomer, non-catastrophic appearance model}
\end{table}

\subsubsection{GRB-tracking catastrophic appearance model}
\begin{table}[H]
\centering
\begin{tabular}{|c||c|c|c|c|c|c|c|c|c|}
\hline Expansion Speed  & $v=.1c$ & $v=.2c$  & $v=.3c$ & $v=.4c$  & $v=.5c$  & $v=.6c$ & $v=.7c$ & $v=.8c$ & $v=.9c$ \\ 
\hline $\alpha$ (appearances/Gly$^3$Gyr) & 1.8 & .22 & .066 & .028 & .014 & .0082 & .0052 & .0035 & .0024 \\ 
\hline $R_1$ (Gly) & .080 & .18 & .30 & .47 & .71 & 1.1 & 1.7 & 3.0 & N/A \\ 	
\hline $EV(obs)$ for unlimited $R$ & 1800 & 230 & 67 & 27 & 13 & 6.7 & 3.5 & 1.8 & .68  \\ 
\hline $EV(obs)$ of non-expanders within $R_1$ & .017 & .024 & .035 & .055 & .091 & .17 & .37 & 1.1 & N/A \\ 
\hline 
\end{tabular} 
\caption{$2 \sigma$ latecomer, GRB-tracking catastrophic appearance model}
\end{table}

\subsubsection{Extreme catastrophic appearance model}
\begin{table}[H]
\centering
\begin{tabular}{|c||c|c|c|c|c|c|c|c|c|}
\hline Expansion Speed  & $v=.1c$ & $v=.2c$  & $v=.3c$ & $v=.4c$  & $v=.5c$  & $v=.6c$ & $v=.7c$ & $v=.8c$ & $v=.9c$ \\ 
\hline $\alpha$ (appearances/Gly$^3$Gyr)  & 13 & 1.6 & .48 & .20 & .10 & .060 & .038 & .025 & .018 \\ 
\hline $R_1$ (Gly) & .044 & .098 & .17 & .26 & .39 & .59 & .92 & 1.6 & N/A  \\ 	
\hline $EV(obs)$ for unlimited $R$ & 2600 & 320 & 92 & 37 & 18 & 9.3 & 4.9 & 2.4 & .95  \\ 
\hline $EV(obs)$ of non-expanders within $R_1$ & .012 & .016 & .024 & .037 & .063 & .12 & .25 & .73 & N/A \\ 
\hline 
\end{tabular} 
\caption{$2 \sigma$ latecomer, extreme catastrophic appearance model}
\end{table}

\end{widetext}

\section{Discussion and Conclusions}

From the model results of the previous section, some conclusions are immediate:

\begin{itemize}
\item There exist large regions of the parameter space that result in $EV(obs) < 1$ for unlimited $R$.  The probability must be regarded as substantial that we can see no expanders or K3 civilizations, no matter how good our observation techniques, even if the saturation of the universe by advanced life is well underway.
\item If humanity has appeared near (or before) the mean time of arrival for civilizations like ours, the prospect of seeing any expanders or isolated K3 civilizations seems poor.  For observation to be likely at the mean arrival time, the limits to technology would have to make intergalactic expansion practical, but not above $\approx .3 c$.  Other authors have concluded that such a barrier could be surpassed even by relatively simple fission rockets~\cite{armstrong2013}, making such low-$v$ scenarios seem less plausible.  If we are living in such a scenario and a positive observation is made, it would most likely be at distances of multiple Gly. 
\item If humanity is a $1 \sigma$ latecomer, observational prospects are better.  Expansion scenarios up to $.7 c$ are likely to be observable, at least in principle.  For such scenarios, $R_1$ remains cosmological and at multiple Gly when expansion above $.4 c$ is practical for maximally advanced life.  
\item In the seemingly unlikely case that humanity is a $2 \sigma$ latecomer, prospects for observation are good in expansion scenarios up to nearly $.9 c$.  In fact, the low-$v$ scenarios of this case are probably ruled out already by existing observations~\cite{griffith2015}, provided they can be regarded as sufficiently thorough searches for K3 civilizations.  Even if we are a $2 \sigma$ latecomer, we still expect to make the first observations at cosmological distances greater than the homogeneity scale (if expansion speeds are greater than $.2 c$), though only expansion speeds above $.6 c$ result in an $R_1$  above $1$ Gly. 
\item In no cases examined do we find expansion scenarios at $.9 c$ such that $R_1$ is defined.  However, if we are a $2 \sigma$ latecomer in the GRB-tracking catastrophic scenario, the probability that zero expanders are visible is $e^{-EV(obs)} \approx 51\%$.  We should not realistically expect to see K3's if intergalactic travel at $v = .9 c$ is practical and the expansion hypothesis is correct, but the probability is not overwhelmingly negative.  The difficulty in observing high-$v$ scenarios comes from two factors -- the very small appearance rate required to satisfy the time of arrival conditions when the expansion speed is high, and the fact that in high-$v$ scenarios, a large fraction of our past light cone (region $C$ of fig. 1) is already known to be devoid of such expanders (else they would already be here).
\item The expected number of non-expanders appearing within $R_1$ is almost always less than the number of expanders that are visible out to $R_1$ (unity), despite the fact that non-expanders had the additional opportunity to appear in region $C$ of fig. 1.  Due to the assumed equality of appearance rates (between expanders and non-expanders), this can be interpreted to mean that we are more likely to observe an expander that came to within $R_1$ from farther away, rather than seeing an expander that appeared within $R_1$ to begin with.  It also means that we expect to have more K3 clusters within visible range than isolated K3 galaxies.
\item Even unrealistically extreme models of galaxy-scale extinction events have a modest effect on our conclusions.  The main effect of such models is to make advanced life nearly impossible in the early universe, before advanced life would have time to arise anyway.  The effect is diminished in more recent times, which are more relevant for the appearance of advanced life.  
\end{itemize}

It is interesting to visualize just how rarely aggressively expanding civilizations arise, according to this analysis.  A typical value for the appearance rate parameter $\alpha$ in a GRB-tracking scenario is of order $10^{-3}$ appearances per Gly$^3$ per Gyr.  In other words, it would take a sphere of radius $\approx 5$ Gly to produce a single aggressive expander in a billion years.  This is a volume encompassing many thousands of superclusters and perhaps a hundred million large galaxies. Similar numbers have been implied by~\cite{fogg1988,armstrong2013}, in their calculation of the number of galaxies that could have reached and colonized the Milky Way.  The great filter~\cite{hanson1998a} implied by this type of universe must be very great indeed. 

We should reflect on our use of the time of arrival distribution as a means of estimating the appearance rate parameter $\alpha$.  This is a form of anthropic reasoning, implicitly utilizing the Self-Sampling Assumption, which exhorts us to reason as if we are a random sample from the set of all comparable observers who will ever have existed~\cite{bostrom2002}.  We have interpreted this to mean that our time of arrival should be typical in the set of human-stage civilizations who will have appeared in a non-K3 galaxy.  Though we have no prior theory to determine $\alpha$, we should feel confident that a scenario in which humanity is a multiple-$\sigma$ latecomer to the universe can be regarded as very improbable without good evidence to the contrary.  This allows us to focus our attention on scenarios where humanity is no more than a $2 \sigma$ latecomer.   

Limitations of our analysis should also be noted.  Since we have assumed a homogeneous universe, the visible geometry of small domains (smaller than the homogeneity scale of the universe) could deviate significantly from the “expanding spheres” assumption.  Similarly, the universe-averaged appearance rate is a rough approximation, at the current level of development.  Taking into account the details of galaxy evolution could presumably make significant changes to $f(t)$, though we have seen that our conclusions seem to remain fairly robust with respect to substantial changes to $f(t)$.  The assumption of behavior uniformity (i.e. a single, constant $\{ v, T \}$ for the expanders) is also debatable -- a mixture of rare fast expanders and common slow expanders, for example, might be expected to change our conclusions substantially, based on our previous calculations~\cite{olson2014}, as would extreme galaxy colonization models that take $T$ to be on the scale of multiple Gyr.  We do not expect the remaining uncertainty in the underlying background cosmological parameters to significantly affect our conclusions, provided that the $\Lambda C D M$ model remains standard.

From the most practical point of view, what does our analysis say?  Our modeling is consistent with the possibility that K3 civilizations and aggressive expanders are present but not observable, or even not present at all.  But we do have a conditional result -- \emph{if} K3 galaxies are observable, then we expect to see them at cosmological distances, as part of a cluster of K3 galaxies whose boundary expands at a middling fraction of the speed of light.  Conditional results of this kind have a curious sort of practicality:  Knowing the location of a streetlight does not tell us where we have dropped our keys, but it does let us know where we have the best chance of finding them.  In this way, such an analysis can be of practical use for future searches of K3 galaxies. 

\bibliography{ref5}{}
\bibliographystyle{plain}

\end{document}